\documentclass[a4paper,fleqn,usenatbib]{mnras}

\usepackage{mathptmx}

\usepackage[T1]{fontenc}
\usepackage{ae,aecompl}

\usepackage{graphicx,epstopdf}	
\usepackage{amsmath}	
\usepackage{amssymb}	
\usepackage{bm}

\title[Hot-Jupiter Roche-lobe Overflow]{Hot-Jupiter Core Mass from Roche-lobe Overflow}

\author[S. Ginzburg and R. Sari]{
Sivan Ginzburg\thanks{E-mail: sivan.ginzburg@mail.huji.ac.il}
and Re'em Sari
\\
Racah Institute of Physics, The Hebrew University, Jerusalem 91904, Israel}

\date{Accepted XXX. Received YYY; in original form ZZZ}

\pubyear{2016}

\begin{document}
\label{firstpage}
\pagerange{\pageref{firstpage}--\pageref{lastpage}}
\maketitle

\begin{abstract}
The orbits of many observed hot Jupiters are decaying rapidly due to tidal interaction, eventually reaching the Roche limit. We analytically study the ensuing coupled mass loss and orbital evolution during the Roche-lobe overflow and find two possible scenarios. Planets with light cores $M_c\lesssim 6M_\oplus$ (assuming a nominal tidal dissipation factor $Q\sim 10^6$ for the host star) are transformed into Neptune-mass gas planets, orbiting at a separation (relative to the stellar radius) $a/R_\star\approx 3.5$. Planets with heavier cores $M_c\gtrsim 6M_\oplus$ plunge rapidly until they are destroyed at the stellar surface. Remnant gas-Neptunes, which are stable to photo-evaporation, are absent from the observations, despite their unique transit radius ($5-10R_\oplus$). This result suggests that $M_c\gtrsim 6M_\oplus$, providing a useful constraint on the poorly-known core mass that may distinguish between different formation theories of gas giants. Alternatively, if one assumes a prior of $M_c\approx 6 M_\oplus$ from the core-accretion theory, our results suggest that $Q$ does not lie in the range $10^6\lesssim Q\lesssim 10^7$. 
\end{abstract}

\begin{keywords}
planets and satellites: composition -- planets and satellites: gaseous planets -- planet-star interactions
\end{keywords}



\section{Introduction}\label{sec:introduction}

One of the major surprises of exoplanet research is the discovery of ``hot Jupiters'' -- giant planets in orbits of a few days \cite[e.g.,][]{MayorQueloz95}. It is yet unclear whether these planets formed in situ \citep{Bodenheimer2000,Ikoma2001,Batygin2016,Boley2016}, migrated inward due to interaction with a gaseous disk \citep{GoldreichTremaine1980,Lin1996}, or are the result of tidal circularization of longer eccentric orbits of scattered planets \citep{RasioFord1996}. 

Regardless of their formation mechanism, the orbits of many of the shortest-period hot Jupiters may shrink significantly within several Gyrs as a result of tidal dissipation in their host stars \citep[e.g.,][]{Rasio1996}. Eventually, these planets are expected to spiral in until they reach the Roche limit and begin to disintegrate. While a complete absorption of the planet by its host star is a possible outcome of the Roche-lobe overflow \citep[e.g.,][]{Jackson2009, Metzger2012}, recent studies suggest that, under some circumstances, hot Jupiters are only partially consumed, leaving behind lower-mass planets \citep{Valsecchi2014,Valsecchi2015,Jackson2016}. Specifically, these studies find that the fate of an overflowing gas giant is dictated by the mass of its rocky core. 

According to the popular core-nucleated accretion theory, gas giants form as gas is gravitationally accreted onto a sufficiently massive rocky (or icy) core \citep{PerriCameron1974,Harris1978,Mizuno1978,Mizuno1980,Stevenson1982}. The critical core mass, around which a gas giant forms, has been the focus of intense theoretical study, with current estimates ranging from about $2M_\oplus$ to $30M_\oplus$, with $M_\oplus$ denoting an Earth mass \citep{BodenheimerPollack86,Pollack96,Ikoma2001,Rafikov2006,Rafikov2011,Lee2014,PisoYoudin2014,LeeChiang2015,Piso2015}. Alternative planet formation theories do not require a core at all \citep{Kuiper1951,Cameron1978,Boss1997,Boss1998,Boss2000}.

On the observational front, measurements of the gravitational field of the solar system's oblate gas giants provides some constraints on their core mass, consistent with the theoretical estimates above 
\citep[see][for a review]{Baraffe2014}. In particular, coreless models are not excluded \citep{Guillot1997, SaumonGuillot2004, FortneyNettelmann2010}.

Here, we utilize the critical role of dense cores in the evolution of overflowing gas giants to constrain the typical core mass of hot Jupiters. Explicitly, we compare the theoretical evolution of observed hot Jupiters that are expected to overflow and lose mass to the observed population of lighter planets (that may be the remnants of incomplete mass transfer), searching for an imprint of the core mass. 

The outline of the paper is as follows. In Section \ref{sec:inspiral} we investigate the hot Jupiters with the closest orbits, which are shrinking rapidly even more due to tidal dissipation. In Section \ref{sec:rlo} we study the ensuing mass transfer, once these planets are sufficiently close to their host stars. In Section \ref{sec:mc} we calculate the critical core mass, which determines the planet's fate, and in Section \ref{sec:remnant} we characterize the surviving low-mass remnants of Roche-lobe overflow. Section \ref{sec:previous} relates this work to previous studies and Section \ref{sec:conclusions} summarizes our main conclusions.  

\section{Inspiralling Hot Jupiters}\label{sec:inspiral}

Tides raised on a host star with radius $R_\star$ and mass $M_\star$ by a planet with mass $M_p$ and separation $a$ cause the planet's orbit to decay on a time-scale
\begin{equation}\label{eq:t_tide}
\begin{split}
t_{\rm tide}\equiv\frac{L}{|\dot{L}|}\sim\frac{a}{|\dot{a}|}&=Qt_{\rm dyn}^\star\left(\frac{a}{R_\star}\right)^{13/2}\frac{M_\star}{M_p}\\&\approx 4\textrm{ Gyr}\left(\frac{a/R_\star}{5}\right)^{13/2}\frac{M_{\rm J}}{M_p},
\end{split}
\end{equation}
calculated by dividing the planet's orbital angular momentum (which is transferred to the stellar spin) $L=M_p\sqrt{GM_\star a}$ by the tidal torque \citep[given by, e.g.,][]{GoldreichSoter1966}.
$t_{\rm dyn}^\star\equiv(G\rho_\star)^{-1/2}\approx 1\textrm{ h}$ and $\rho_\star$ denote the star's  dynamical time and density, respectively, $G$ is Newton's constant, and $M_{\rm J}$ is Jupiter's mass. The numerical estimate is for a solar-like star and adopting a frequency-independent stellar tidal dissipation factor $Q\sim 10^6$ 
\citep[see][and references therein]{EssickWeinberg2016}. Tides raised on the planet keep it tidally locked, since a hot Jupiter's synchronization time-scale is significantly shorter than $t_{\rm tide}$ above \citep{Guillot1996}.

\begin{figure}
	\includegraphics[width=\columnwidth]{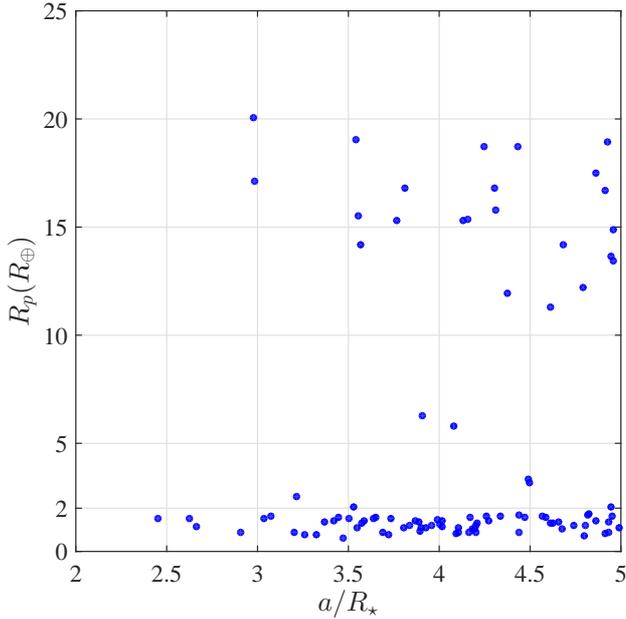}
	\caption{Observed planets from the exoplanets.org database \citep{Han2014} with separations (in units of the stellar radius) $a/R_\star<5$. Planets with radii $R_p>10R_\oplus$ are Jupiters that are expected to spiral inward due to tidal dissipation within a few Gyrs, according to Equation \eqref{eq:t_tide}. Error bars are typically of the order of 10\% and are not displayed for clarity.}
	\label{fig:ator}
\end{figure}

In Fig. \ref{fig:ator} we present observed planets with $a/R_\star<5$. All the planets with radii $R_p>10R_\oplus$ have masses $M_p\gtrsim M_{\rm J}$ and will therefore spiral inward within their host star's lifetime (the hosts are main-sequence stars), according to Equation \eqref{eq:t_tide}. The discovery of several Jupiters with $a/R_\star$ as low as 3, implying $t_{\rm tide}\ll\textrm{Gyr}$, has led some authors to postulate a larger $Q$ \citep{OgilvieLin2007,Hellier2009,PenevSasselov2011}. Others remark, on the other hand, that dynamical scattering of Jupiters into short orbits (see Section \ref{sec:introduction}) may repopulate these rapidly decaying orbits \citep[e.g.,][]{EssickWeinberg2016}. Additionally, a sample of hundreds of observed hot Jupiters would inevitably harbour a few short-lived planets, since Equation \eqref{eq:t_tide} predicts a detection probability of $\sim 1\%$ for $a/R_\star=3$.

We conclude that there is an observed population of over a dozen hot Jupiters that are inspiralling with a time-scale $t_{\rm tide}\lesssim\textrm{Gyr}$ \cite[see also][]{EssickWeinberg2016}. Their typical mass is $\approx 1.4M_{\rm J}$ and the typical radius is $\approx 1.4R_{\rm J}$. By coincidence, the typical stellar and planetary densities are similar $\rho_\star\approx\rho_p\approx 0.6\textrm{ g cm}^{-3}$. The low planetary density (relative to Jupiter) is a manifestation of the extensively-studied hot-Jupiter inflation puzzle \citep[see, e.g.,][and references therein]{SpiegelBurrows2013, GinzburgSari2015, GinzburgSari2016}, which is also evident in Fig. \ref{fig:ator}. The low stellar density (relative to the Sun) reflects the typically higher mass of the host stars\footnote{$M_\star\approx 1.3M_\odot$ also marks the transition between stars with convective and radiative envelopes \citep{Kraft67,VSP2013}. Some studies suggest that stars with radiative envelopes have a significantly higher $Q$ \citep{Zahn1977,VanEylen2016}. Considering this effect would roughly halve the number of inspiralling planets in Fig. \ref{fig:ator}.} $M_\star\approx 1.3 M_\odot$ \citep[see][for a mass-radius relation]{Torres2010}.
In the following sections we study the fate of such planets as they spiral towards their host stars.

\section{Stable Roche-lobe Overflow}\label{sec:rlo}

As a planet inspirals towards its host star, it reaches the Roche limit at a separation
\begin{equation}\label{eq:roche}
\frac{a}{R_\star}\simeq 2.4\left(\frac{\rho_\star}{\rho_p}\right)^{1/3},
\end{equation}
where the numerical coefficient is suitable for an incompressible planet and it changes to $\simeq 2.0$ for compressible bodies \citep[e.g.][and references therein]{Rappaport2013}. Based on the interpolation between the two regimes in \citet{Rappaport2013} and Jupiter's central to mean density ratio \citep[e.g.][]{Peebles1964,ChabrierBaraffe2000}, Equation \eqref{eq:roche} is accurate to within a few percent. As the planet loses mass through Roche-lobe overflow (RLO), the approximation improves since the planet becomes even less compressible (see Section \ref{sec:compress}).

Once the planet reaches the Roche limit, stable mass transfer may ensue, during which the planet remains on the verge of the Roche limit, given by Equation \eqref{eq:roche}. The mass-loss and orbital evolution are driven by tidal angular momentum transfer, with a time-scale $\sim t_{\rm tide}$, given by Equation \eqref{eq:t_tide}. See Appendix \ref{sec:stability} for a derivation and a discussion. See \citet{Metzger2012} for possibly observable transients in case the Roche limit lies within $R_\star$ or if the mass transfer is unstable. 

\subsection{Gas Compressibility}\label{sec:compress}

According to Equation \eqref{eq:roche}, the planet's separation during RLO is simply a measure of its density $a\propto\rho_p^{-1/3}$. In this section we calculate $\rho_p$ as a function of the planet's mass.

We first consider a zero-temperature coreless gas giant. The pressure inside such a planet is given by a simple equation of state \citep[see, e.g.,][]{Padmanabhan2001,ArrasBildsten2006,GinzburgSari2015}
\begin{equation}\label{eq:eos}
	P=K\rho_p^{5/3}\left[1-\left(\frac{\rho_0}{\rho_p}\right)^{1/3}\right]=\beta GM_p^{2/3}\rho_p^{4/3},
\end{equation}
where the first term in brackets is due to electron degeneracy, with $K\sim h^2m_e^{-1}m_p^{-5/3}$ ($h$ is the Planck constant and $m_e$ and $m_p$ are the electron and proton masses), and the second term is due to electrostatic forces ($\rho_0$ is the zero-pressure density of cold gas). This pressure balances the one exerted by gravity $\sim GM_p^2/R_p^4$, which appears on the right hand side of Equation \eqref{eq:eos} with an order of unity constant $\beta$. We rewrite Equation \eqref{eq:eos} as
\begin{equation}\label{eq:rho_cold}
	\left(\frac{\rho_p}{\rho_0}\right)^{1/3}=1+\left(\frac{M_p}{M_{\rm max}}\right)^{2/3},
\end{equation}
with $M_{\rm max}\equiv\beta^{-3/2}(K/G)^{3/2}\rho_0^{1/2}$. It is easy to see from Equation \eqref{eq:rho_cold} that the planet's radius $R_p\propto(M_p/\rho_p)^{1/3}\propto M_p^{1/3}/[1+(M_p/M_{\rm max})^{2/3}]$ peaks at $M_p=M_{\rm max}$, with $R_p\propto M_p^{1/3}$ for $M_p\ll M_{\rm max}$ and $R_p\propto M_p^{-1/3}$ for $M_p\gg M_{\rm max}$. We can easily calibrate $M_{\rm max}\simeq 3M_{\rm J}$ from the peak in the radius-mass curves calculated in previous studies \citep{ZapolskySalpeter1969, Fortney2007}.
Equation \eqref{eq:rho_cold} demonstrates that planets with a mass $M_p\ll M_{\rm max}$ are incompressible. 

Next, we turn our attention to the low densities of observed hot Jupiters in comparison with cold gas giants (see Section \ref{sec:inspiral}). These low densities indicate that hot Jupiters retain high interior temperatures \citep[e.g.][]{ArrasBildsten2006}, and may therefore deviate from the zero-temperature $\rho_p(M_p)$ relation of Equation \eqref{eq:rho_cold}. Specifically, the deviation in density is given by $d\rho/\rho=f(\theta)d\theta$, with $\theta$ denoting the degeneracy parameter, defined as the ratio of the ions' thermal pressure to the electrons' degeneracy pressure, and $f=-3/(1+\theta)$ \citep{GinzburgSari2015}. Consequently, the ratio of hot to cold densities $\rho(\theta)/\rho(\theta=0)$ is a function of $\theta$ alone. \citet{Guillot2005} finds that $\theta$ is roughly uniform across a hot Jupiter's interior profile, because the pressure of both thermal ions and degenerate electrons scales as $P\propto\rho^{5/3}$. 
As a result, when the outer layers of a gas giant are removed through Roche-lobe overflow, the remaining planet retains its original $\theta$, assuming the mass-loss is adiabatic \citep[the thermal evolution time-scale of the planet's interior is longer than the orbital evolution for tidally evolving planets, i.e., below the dashed black line in Fig. \ref{fig:evolution} and \ref{fig:evolution_mc}; see][]{WuLithwick2013,GinzburgSari2016}.

We conclude that $\theta$, and therefore the relative inflation 
$\rho_p(\theta)/\rho_p(\theta=0)$, is conserved during RLO. This result allows us to generalize Equation \eqref{eq:rho_cold}, which was derived for cold planets ($\theta=0$), to hot (inflated) planets ($\theta>0$), as they lose mass:
\begin{equation}\label{eq:rho_p_hot}
	\rho_p^{1/3}\propto 1+\left(\frac{M_p}{M_{\rm max}}\right)^{2/3}.
\end{equation}
Notice that the density in Equation \eqref{eq:rho_p_hot} is calibrated to a lower value (i.e. lower than $\rho_0$; see Section \ref{sec:mc} for a detailed calibration) in comparison with Equation \eqref{eq:rho_cold}, due to the inflation.

\subsection{Asymptotic Trajectories}\label{sec:asymptotic}

\begin{figure}
	\includegraphics[width=\columnwidth]{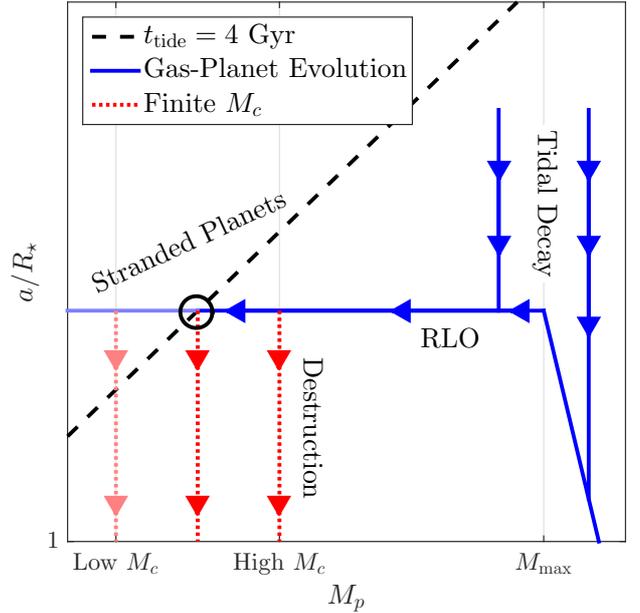}
	\caption{Schematic evolution of short-period hot Jupiters. Gas giants in short orbits inspiral towards the Roche limit at $a/R_\star\approx 2.4$, given by Equation \eqref{eq:roche} with typical stellar and planetary densities. The ensuing Roche-lobe overflow (RLO) is at an approximately constant density for $M_c<M_p<M_{\rm max}$, according to Equation \eqref{eq:rho_p_hot}. When the planet is reduced to $M_p\sim M_c$, the density increases and the planet plunges towards the star (dotted red lines for three values of the core mass $M_c$). Planets with massive cores (rightmost dotted red line) are destroyed within a few Gyrs. Planets with lighter cores (pale lines) are stranded on the dashed black line, which marks the evolution time and is given by Equation \eqref{eq:t_tide}.}
	\label{fig:evolution}
\end{figure}

The stages of orbital evolution and mass loss are depicted schematically in Fig. \ref{fig:evolution} by considering the asymptotic behaviour of Equation \eqref{eq:rho_p_hot}. First, the planet's orbit decays until it reaches the Roche limit, at a separation $a/R_\star\approx 2.4$, according to Equation \eqref{eq:roche}, and substituting typical stellar and planetary densities (see Section \ref{sec:inspiral}). 
Next, the planet loses mass though RLO and its orbit evolves as $a\propto\rho_p^{-1/3}$. According to Equation \eqref{eq:rho_p_hot}, the density and therefore the separation of gas planets are approximately constant for $M_p\ll M_{\rm max}$.
When the planet's mass becomes comparable to the mass of its dense core $M_p\sim M_c$, its mean density increases rapidly (see Section \ref{sec:dense_core} for a more accurate analysis) as the atmosphere is lost, until $\rho_p=\rho_c$, with $\rho_c\approx 6\textrm{ g cm}^{-3}(M_c/M_\oplus)^{1/4}$ marking the rocky core's density \citep[e.g.][]{Valencia2006}. According to Equation \eqref{eq:roche}, the sudden increase in density causes the planet to plunge towards the star until $a/R_\star\simeq 1$ and the planet is destroyed at the stellar surface. 

As seen in Fig. \ref{fig:evolution}, planets with massive cores (rightmost dotted red line) are destroyed within a few Gyrs. Planets with lighter cores (leftmost dotted red line), or without a core (solid blue line), on the other hand, are stranded in a slow stable RLO (marked by a black circle). The gas mass fraction of these stranded planets is $M_{\rm atm}/M_p\gtrsim 0.5$, because the descent towards the star begins when $M_p$ is comparable to $M_c$, as explained above. By combining Equations \eqref{eq:t_tide} and \eqref{eq:roche} and substituting typical $\rho_p$ and $\rho_\star$ (see Section \ref{sec:inspiral}) we estimate that the critical core mass (middle dotted red line) that distinguishes the two cases is a few $M_\oplus$.

The simple analysis above demonstrates how the typical hot-Jupiter core mass $M_c$ can be inferred from the outcome of Roche-lobe overflow.
If $M_c$ is low then we expect to observe a population of gas rich ($\gtrsim 50\%$ in mass) super Earths at a separation $a/R_\star\approx 2.4$. Such a population seems to be absent from the observations (see Section \ref{sec:remnant}), implying that $M_c$ is high. However, from Equations \eqref{eq:t_tide} and \eqref{eq:roche}, the critical core mass is a sensitive function of the planet's density during RLO $M_c\propto\rho_p^{-13/6}$, necessitating a more accurate calculation.
    
\section{Critical Core Mass}\label{sec:mc}

In Section \ref{sec:asymptotic} we identified a critical core mass $M_c$, which determines the outcomes of Roche-lobe overflow. Planets with heavier cores are destroyed by the star within a few Gyrs, while planets with lighter cores survive as short-period gas-rich super Earths. In this section we calculate the critical $M_c$ more accurately. Explicitly, according to Equation \eqref{eq:roche}, the planet's trajectory in the $a$-$M_p$ plane is simply a function of the $\rho_p(M_p)$ relation. In Fig. \ref{fig:evolution} we approximated the density as constant for $M_c<M_p<M_{\rm max}$, followed by a sharp increase for $M_p\simeq M_c$, resulting in simple trajectories. Here, we apply a realistic $\rho_p(M_p)$ relation, which takes into account the compressibility of the gas and the transition from a gas-dominated to a core-dominated planet, leading to more complex trajectories.

Fig. \ref{fig:evolution_mc} displays the evolution of gas giants with realistic densities. The solid blue line, which is adequate for a coreless planet, is according to Equation \eqref{eq:rho_p_hot}, calibrated to a typical inspiralling hot Jupiter (i.e., using the coincidence $\rho_p=\rho_\star$ for $M_p=1.4M_{\rm J}$; see Section \ref{sec:inspiral}). At the onset of RLO $M_p\approx M_{\rm J}$ is comparable to $M_{\rm max}\simeq 3M_{\rm J}$. Therefore, as illustrated in Fig. \ref{fig:evolution_mc}, mass loss during RLO decompresses the planet according to Equation \eqref{eq:rho_p_hot}, causing its orbit to expand in consequence of Equation \eqref{eq:roche}.
    
\subsection{Dense Core}\label{sec:dense_core}

\begin{figure}
	\includegraphics[width=\columnwidth]{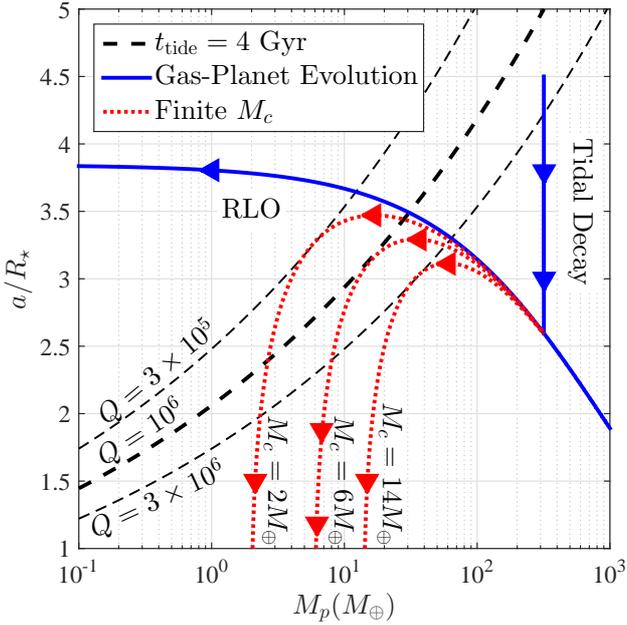}
	\caption{Same as Fig. \ref{fig:evolution}, but with realistic densities $\rho_p(M_p)$, given by Equation \eqref{eq:rho_final}, and inserted into Equation \eqref{eq:roche}. The orbital evolution is plotted for core masses $M_c/M_\oplus=2,6,14$ (dotted red lines, left to right) and for a coreless planet (solid blue line). During RLO, the density first decreases as gravitational pressure is lifted and then increases as the core becomes the dominant component. The minimum density is translated to a maximum separation, since $a\propto\rho_p^{-1/3}$. Planets with $M_c>6M_\oplus$ are destroyed in a few Gyrs, assuming a nominal stellar tidal dissipation factor $Q=10^6$ (thick middle dashed black line), while planets with lighter cores survive as ``gas Neptunes'' with $15M_\oplus<M_p<30M_\oplus$ and $M_{\rm atm}/M_p>60\%$. The thin dashed black lines mark $Q=3\times 10^5$ and $Q=3\times 10^6$ (left to right) which correspond to critical core masses of $M_c/M_\oplus=2$ and 14.} 
	\label{fig:evolution_mc}
\end{figure}

Finally, we consider planets with cores of a finite mass $M_c$ and density $\rho_c$, engulfed by a gas atmosphere with a mass $M_{\rm atm}\equiv M_p-M_c$ and density $\rho_{\rm atm}$.
The mean density $\rho_p$ of such a two-layer planet is given by dividing $M_p$ by the total volume, $M_c/\rho_c+M_{\rm atm}/\rho_{\rm atm}$: 
\begin{equation}\label{eq:rho_core}
\frac{\rho_p}{\rho_{\rm atm}}=\frac{M_p/M_c}{\rho_{\rm atm}/\rho_c+M_{\rm atm}/M_c}\approx\frac{M_p}{M_{\rm atm}}.
\end{equation}
We consider envelopes that are massive enough to dominate the planet's volume $M_{\rm atm}/M_c\gg\rho_{\rm atm}/\rho_c$, and justify this assumption below. We see from Equation \eqref{eq:rho_core} that, for a roughly constant $\rho_{\rm atm}$, the planet's density $\rho_p\propto M_p/(M_p-M_c)$ increases significantly when $M_p\sim M_c$, motivating the sharp increase in density there, used in our approximate analysis in Section \ref{sec:asymptotic}. However, as illustrated in Fig. \ref{fig:evolution_mc} (dotted red lines), the gradual density increase implied by Equation \eqref{eq:rho_core}  is translated to a gradual shrinkage of the orbit by Equation \eqref{eq:roche}. 
This finer estimate of the density allows us to calculate the orbital evolution more accurately and to characterize the remnants of RLO (if there are any, see Section \ref{sec:rlo} and the discussion below) better.

To complete our realistic $\rho_p(M_p)$ relation, we substitute into Equation \eqref{eq:rho_core} $\rho_{\rm atm}(M_{\rm atm})$ from Equation \eqref{eq:rho_p_hot}, which describes the atmosphere's density for $M_{\rm atm}\gtrsim M_c$. Although Equation \eqref{eq:rho_p_hot} was derived for gaseous planets, it is valid for $M_{\rm atm}\lesssim M_c$ as well. The reason is that, since $M_c\ll M_{\rm max}$, $\rho_{\rm atm}$ is almost constant for $M_{\rm atm}\sim M_c$, according to Equation \eqref{eq:rho_p_hot}. For $M_{\rm atm}\lesssim M_c$ it is clear that the atmosphere is even less compressible (due to the weaker gravitational pressure), so $\rho_{\rm atm}$ is constant in that regime, as in Equation \eqref{eq:rho_p_hot}. With this substitution, our density-mass relation is
\begin{equation}\label{eq:rho_final}
\rho_p^{1/3}\propto\left[1+\left(\frac{M_{\rm atm}}{M_{\rm max}}\right)^{2/3}\right]\left(\frac{M_p}{M_{\rm atm}}\right)^{1/3},
\end{equation}
which is again calibrated to $\rho_p=\rho_\star$ for $M_p=1.4M_{\rm J}$ (this coincidence is explained in Section \ref{sec:inspiral}). The first term in Equation \eqref{eq:rho_final} accounts for the compression of the gas atmosphere for $M_{\rm atm}\gtrsim M_{\rm max}\gg M_c$ and approaches the zero-pressure density of the gas for $M_{\rm atm}\lesssim M_c\ll M_{\rm max}$ (as explained above, the exact form of this term is unimportant for $M_{\rm atm}\lesssim M_c$). The second term describes the increase of the mean density, as the core becomes the dominant component, when $M_{\rm atm}\lesssim M_c$. Essentially, Equation \eqref{eq:rho_final} is a trivial interpolation of these two regimes, which are uncoupled thanks to $M_c\ll M_{\rm max}$. For a coreless planet $M_c=0$, Equation \eqref{eq:rho_p_hot} is reproduced. 

Fig. \ref{fig:evolution_mc} displays the orbital evolution of overflowing gas giants with densities given by Equation \eqref{eq:rho_final}. The evolution is characterized by an orbital expansion that precedes the planet's final inspiral towards the star (see the caption of Fig. \ref{fig:evolution_mc} for details). A similar behaviour was found in previous studies \citep{Valsecchi2014,Valsecchi2015,Jackson2016}, and it is intuitively explained by the simpler Fig. \ref{fig:evolution}. Quantitatively, our estimate of the critical core mass, above which planets are destroyed within a few Gyrs, is $M_c=6M_\oplus$ (see Fig. \ref{fig:evolution_mc}). The properties of remnant planets (if $M_c<6M_\oplus$) are described in Section \ref{sec:remnant}.

\subsection{Sensitivity to $Q$}\label{sec:q}

The approximate analysis in Section \ref{sec:asymptotic}, combined with Equation \eqref{eq:t_tide}, yields a linear dependence of the critical core mass on the stellar tidal dissipation parameter $M_c\propto Q$. Fig. \ref{fig:evolution_mc} illustrates that this linear relation is valid only approximately in the more elaborate calculation that we preformed in this section, with $M_c/M_\oplus=2$ and 14 for $Q=3\times 10^5$ and $3\times 10^6$, respectively. 

Notice that for high $Q$ values, not all the hot Jupiters in Fig. \ref{fig:ator} will inspiral within a few Gyrs, as seen in Fig. \ref{fig:evolution_mc}. For low $Q$ values, the remnant planets are less massive and their atmospheres are therefore susceptible to photo-evaporation (see Section \ref{sec:remnant}). 

\section{Remnant Planets}\label{sec:remnant}

In Section \ref{sec:rlo} we demonstrated that hot Jupiters with light cores experience incomplete Roche-lobe overflow and survive as gas-rich lower mass planets. Using the results of Section \ref{sec:mc}, we can now characterize these surviving planets more accurately.

Fig. \ref{fig:evolution_mc} shows that if $M_c<6M_\oplus$, the planet's evolution slows down, and it is stranded in RLO instead of plunging towards the star. Concretely, such planets appear in Fig. \ref{fig:evolution_mc} between the solid blue line (no core) and the middle dotted red line ($M_c=6M_\oplus$), with the leftmost dotted red line ($M_c=2M_\oplus$) as an example. As seen in the figure, after a few Gyrs (i.e. on the thick middle dashed black line), these surviving planets constitute a very distinct population. Explicitly, all of them are expected to be ``gas Neptunes'' with masses $15M_\oplus<M_p<30M_\oplus$ and $M_{\rm atm}/M_p>60\%$ (since $M_c<6M_\oplus$), stranded at $a/R_\star\approx 3.5$. The narrow range of remnant masses and separations is qualitatively explained by the simpler analysis in Section \ref{sec:asymptotic}. The high gas fraction justifies our approximation $M_{\rm atm}/M_c\gg\rho_{\rm atm}/\rho_c\approx 3\%$ (see Section \ref{sec:dense_core}). According to Equation \eqref{eq:rho_core}, this approximation breaks down when $\rho_p/\rho_{\rm atm}\sim (M_p/M_c)/(\rho_{\rm atm}/\rho_c)\approx \rho_c/\rho_{\rm atm}$, i.e. when $\rho_p\approx\rho_c$, affecting only the lower parts of the dotted red curves in Fig. \ref{fig:evolution_mc} (at $a/R_\star\simeq 1$), which do not influence our results.

In addition to RLO, close-in planets lose mass through photo-evaporation by high-energy stellar radiation. While hot Jupiters lose only $\sim 1\%$ of their atmosphere during their lifetime \citep{MurrayClay2009}, lighter super Earths can be stripped of their gas envelopes entirely \citep{Lopez2012}. By considering the high stellar flux $\propto(a/R_\star)^{-2}$ and the intermediate mass of our remnant planets, their atmospheres should apparently be marginally unstable to evaporation \citep[see Fig. 8 of][but with $M_p$ instead of $M_c$, since $M_{\rm atm}\gtrsim M_c$]{LopezFortney2013}. However, since the stellar UV activity decreases rapidly afterwards, almost all of the mass loss occurs during the star's first $\sim 100\textrm{ Myr}$ \citep{Lopez2012,LopezFortney2013,OwenWu2013}. The remnant planets that we consider here, on the other hand, reach their low mass only after Gyrs of tidal decay (as Jupiters, see Section \ref{sec:inspiral}). At this stage, the high-energy flux is too low to erode their atmospheres significantly \citep[see Fig. 1 of][]{LopezFortney2013}. We conclude that the remnant gas Neptunes retain their heavy atmospheres despite the high incident stellar flux that they receive (see also the discussion in Section \ref{sec:previous}).

Even without detailed modelling of the structure and thermal evolution (after a planet is stranded, its thermal evolution may become faster than its tidal evolution, so it is no longer adiabatic) of gas Neptunes, it is clear that their radii are expected to be in the range $5-10 R_\oplus$ \citep{ZapolskySalpeter1969,Fortney2007,Lopez2012}. This range is remarkably empty, as seen in Fig. \ref{fig:ator}. Moreover, one of the two planets in the range appears to be a disintegrating sub-Earth \citep{Rappaport2012}; the other orbits an exceptionally low-density star \citep{Morton2016}, implying, according to Equation \eqref{eq:roche}, a significantly smaller separation, if it were an RLO remnant. 

In summary, gas Neptunes, which are the expected remnants of incomplete RLO of hot Jupiters, are absent from the observations. We therefore come to the conclusion that the majority of hot-Jupiter cores have a mass $M_c>6M_\oplus$. In this case, the planets complete the RLO and do not leave a remnant, as seen in Fig \ref{fig:evolution_mc}. Although a statistical analysis of the chance to observe non-stranded ($M_c>6 M_\oplus$) overflowing planets \citep[analogous to the analysis in][for inspiralling Jupiters]{Penev2012} is beyond the scope of this work, it is clear that such an analysis would only increase our lower limit on $M_c$. The reason is that even planets with cores somewhat heavier than $6M_\oplus$ spend a long time close to the thick black dashed line in Fig. \ref{fig:evolution_mc}, where the tidal evolution time is long, and there is a non-negligible chance to observe them, yet none are observed. Finding a large population of close-orbiting gas Neptunes would have reversed our conclusion, though additional formation channels (rather than RLO of hot Jupiters) should also be considered in this case.

\section{Relation to Previous Works}\label{sec:previous}

In this section we outline the main differences between this work and previous studies of the Roche-lobe overflow of hot Jupiters.

First and foremost, the goal of our study is different. Previous works \citep{Valsecchi2014,Valsecchi2015,Jackson2016} studied the correlations between the mass and the period of mainly-rocky super Earths and checked whether these could be the surviving cores of hot Jupiters that underwent RLO. Here, we argue that none of the observed super Earths were produced by RLO. Instead, we constrain the hot-Jupiters' core mass by the absence of more massive low-density ``gas Neptunes'' (see Section \ref{sec:remnant}).

Another major difference is our method. Previous studies employed rather complex mass-radius relations and calculated the orbital evolution numerically using the MESA code. We, on the other hand, address the problem with a simpler analytical approach, allowing us to intuitively draw qualitative and quantitative conclusions and to easily test the parameter space (see Section \ref{sec:q}). We did not introduce any severe approximations, so our analytical method does not restrict our accuracy. 

\begin{figure}
	\includegraphics[width=\columnwidth]{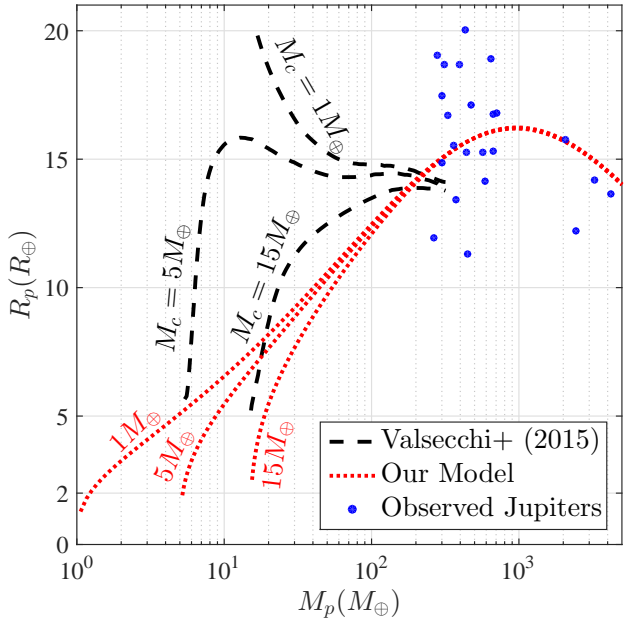}
	\caption{Mass-radius relation for overflowing Jupiters with different core masses ($M_c/M_\oplus=1,5,15$, as labelled). The population of observed inspiralling Jupiters (blue dots) is from Fig. \ref{fig:ator}. The dashed black lines are from Figure 7 of \citet{Valsecchi2015}, while the dotted red lines are according to Equation \eqref{eq:rho_final}. The contrast between the two models is explained by two main differences: (i) We account for inflation due to deep heating, which is absent from \citet{Valsecchi2015}. (ii) In \citet{Valsecchi2015}, planets that are reduced to a low mass are re-inflated (reheated) by the stellar irradiation, while we assume that the overflow is adiabatic \citep[following][]{WuLithwick2013,GinzburgSari2016}. Our models are truncated when the approximation $M_{\rm atm}/M_c\gg \rho_{\rm atm}/\rho_c$ breaks down (see Section \ref{sec:dense_core}).}
	\label{fig:mass_rad}
\end{figure}
   
Putting these differences aside, many of our results qualitatively reproduce previous conclusions. For instance, both \citet{Valsecchi2014} and \citet{Jackson2016} notice that there are no gas-rich planets currently at RLO, leading \citet{Valsecchi2014} to conclude that most hot Jupiters have rocky cores. Here, we take this conclusion one step further and constrain the core mass more accurately. Similarly, \citet{Valsecchi2015} and \citet{Jackson2016} find that planets may be stranded near a maximum period that is anti-correlated with their core mass, as in Fig. \ref{fig:evolution_mc}. Nonetheless, we list below several critical quantitative differences:

\begin{enumerate}
\item
{\it Mass-radius relation}. The previous studies refer to a variety of mass-radius relations for irradiated planets, taken from earlier numerical planetary evolution models \citep{Fortney2007,BatyginStevenson2013,LopezFortney2014}. However, all of these models evolve planets at a constant $M_p$ for a few Gyrs. The cooling of lighter planets is hindered more by stellar irradiation; hence the degeneracy parameter $\theta$ (see Section \ref{sec:compress}) and the relative inflation scale inversely with $M_p$ in the models \citep[see][]{WuLithwick2013,GinzburgSari2016}. Overflowing planets, on the other hand, retain their original small $\theta$, which corresponds to a few Gyrs evolution of the original Jupiter mass, as explained in Section \ref{sec:compress}. As a result, using the models above flattens the $R_p(M_p)$ relation and steepens the $\rho_p(M_p)$ curve for $M_p\lesssim M_{\rm J}$. Consequently, using Equation \eqref{eq:roche}, \citet{Valsecchi2014}, who incorporate the models of \citet{Fortney2007} and \citet{BatyginStevenson2013},  
overestimate the orbital expansion at the onset of RLO (see Fig. \ref{fig:evolution_mc}). More recent studies \citep{Valsecchi2015,Jackson2016} calculate the mass-radius relation during RLO directly, using MESA. In Fig. \ref{fig:mass_rad} we compare our mass-radius relation to \cite{Valsecchi2015}. The contrast between the two models is explained by two main differences. First, our model accounts for inflation due to deep heating \citep[such as Ohmic dissipation, see][]{BatyginStevenson2010}, which is absent from \citet{Valsecchi2015}, despite being necessary in order to explain the large radii of observed hot Jupiters. Second, the overflowing planets in \citet{Valsecchi2015} are re-inflated (reheated) to low densities by the stellar irradiation as they lose mass, contrary to numerical
\citep{WuLithwick2013} and analytical \citep{GinzburgSari2016} studies that find prohibitively long re-inflation time-scales. The origin of the discrepancy (for $M_p\sim M_J$) between \cite{Valsecchi2015} and \cite{WuLithwick2013}, who also use MESA, is unclear, nor is it explained in \cite{Valsecchi2015}. Our analytical estimates \citep{GinzburgSari2016} agree better with \citet{WuLithwick2013}. However, since both \citet{WuLithwick2013} and \citet{GinzburgSari2016} focus on roughly Jupiter-mass planets, the re-inflation time for significantly lower masses might be shorter. Another possibility is that $\theta$ is not conserved even during adiabatic mass-loss due to deviations from the $P\propto\rho^{5/3}$ scaling (see Section \ref{sec:compress}). From Fig. \ref{fig:evolution_mc}, such re-inflation would only increase our lower bound on $M_c$ by increasing the separation at low $M_p$ values.
The low densities in \citet{Valsecchi2015} result in an overestimate of the orbital expansion during RLO \citep[see also][]{Jackson2016}. Moreover, the increase in $R_p$ as the planet loses mass (see Fig. \ref{fig:mass_rad}) has a destabilizing  effect, leading \cite{Valsecchi2015} to consider unstable mass-transfer, while we find that the mass transfer is stable (see Appendix \ref{sec:stability}).
\item
{\it Photo-evaporation}. As explained in Section \ref{sec:remnant}, a combination of the high mass ($M_p>15M_\oplus$) of remnant planets and the late stage of stellar evolution ($\gg 100\textrm{ Myr}$), in which they are reduced to this mass, renders photo-evaporation (PE) irrelevant for Roche-lobe overflow. While \citet{Valsecchi2014} and \citet{Jackson2016} ignore PE as well, \citet{Valsecchi2015} incorporate it in their MESA runs. They find that PE removes a significant portion of a remnant gas-Neptune's atmosphere in a few Gyrs. According to Section \ref{sec:remnant}, this result is at odds with the results of \citet{LopezFortney2013}, considering the planet's separation, mass, and age. Using the results of \citet{LopezFortney2013}, we estimate that only a minor part of a gas Neptune's mass (no more than a few $M_\oplus$) is removed by PE.
In fact, this difference is also evident in Fig. 2 of \citet{Valsecchi2015}, which compares their PE prescription to \citet{LopezFortney2013}. Despite apparent similarities, the figure reveals that the prescription of \citet{Valsecchi2015} removes $\approx 3$ times the mass, in comparison with \citet{LopezFortney2013}, at late times ($\gtrsim\textrm{Gyr}$) which are relevant for our scenario. This modest difference is enough to explain the dissimilar end-results of PE (gas Neptunes vs. rocky super-Earths with $M_{\rm atm}\ll M_c$).
\end{enumerate}

\section{Conclusions}\label{sec:conclusions}

More than a dozen observed hot Jupiters will inevitably inspiral towards their host stars within a few gigayears, due to tidal interaction (see Fig. \ref{fig:ator}). At some stage, they will reach the Roche limit and start losing mass via Roche-lobe overflow (RLO). Capitalizing on previous works, these doomed planets can serve as a laboratory to study the structure of hot Jupiters. In particular, their evolution during RLO critically depends on their core mass $M_c$ \citep{Valsecchi2014,Valsecchi2015,Jackson2016}.

Here, we characterized this population of inspiralling Jupiters and affirmed that their mass overflow is stable, with a rate determined by the tidal evolution time-scale $t_{\rm tide}$. We identified two possible RLO outcomes:
\begin{enumerate}
\item\label{item:low_mass}
$M_c<6M_\oplus$: The planet's mass decreases until $t_{\rm tide}\gtrsim\textrm{Gyr}$, stranding the remnant planet (which is reduced to a mass $15M_\oplus<M_p<30M_\oplus$) at a separation $a/R_\star\approx 3.5$.
\item\label{item:high_mass}
$M_c>6M_\oplus$: The planet's atmosphere is gradually removed, until the dense core increases the planet's mean density $\rho_p$. During RLO, $a\propto\rho_p^{-1/3}$, causing the orbit to shrink until the planet quickly crushes on the stellar surface.  
\end{enumerate} 

The remnants in option \ref{item:low_mass} should have been Neptune mass planets, composed primarily of gas. Owing to a combination of their relatively high mass and the late stage in stellar evolution at which they are reduced to this mass, these ``gas Neptunes'' are not affected by photo-evaporation, in contrast to lighter super Earths \citep{Lopez2012,LopezFortney2013,OwenWu2013}.
Surviving gas Neptunes would have a distinct imprint on the observations as $5-10R_\oplus$ planets at $a/R_\star\approx 3.5$. However, such planets are remarkably absent from the transiting population (see Fig. \ref{fig:ator}).

The lack of short-period gas Neptunes, compared with the relative abundance of their possible progenitors (inspiralling Jupiters) leads us to the conclusion that option \ref{item:high_mass} is the prevalent scenario, i.e., the typical hot-Jupiter core mass satisfies $M_c>6M_\oplus$. This constraint is similar to modern theoretical estimates of the minimum $M_c$ required for Jupiter formation by the core-accretion scenario \citep{LeeChiang2015,Piso2015}. However, observations and alternative formation theories do not rule out lighter cores or coreless planets \citep{Boss2000,Baraffe2014}. Therefore, the indirect observational evidence that is presented in this work provides a useful lower limit on the core mass, which may help to establish the core-accretion theory.  

We employ a simple mass-density relation, enabling us to solve the planet's orbital evolution analytically. Despite its simplicity, our analytical $\rho_p(M_p)$ function, given by Equation \eqref{eq:rho_final}, fully incorporates the two-layer (core and gas) structure of the planet, the compressibility of these layers, and the inflation due to stellar irradiation. Accuracy is achieved by a simple calibration to numerical mass-radius relations and to the observations. Using this density function, we reproduced the complex trajectories of overflowing planets in the $a-M_p$ plane. Specifically, planets first decompress as the gravitational pressure is lifted, and then increase in mean density as their dense core becomes the dominant component. Since $a\propto\rho_p^{-1/3}$, this minimum in $\rho_p$ is translated to a maximum separation \citep[see also][]{Valsecchi2014,Valsecchi2015,Jackson2016}. Our analytical approach allows us to easily test our assumptions and sensitivity to the various parameters.       

Several aspects of our model deserve further attention, which we leave to future work:
\begin{enumerate}
\item
The results depend on the poorly-known stellar tidal dissipation parameter $Q$ (see Section \ref{sec:q}). A more accurate $Q$ value would secure our conclusions. Alternatively, given an estimate of $M_c$ (e.g. from core-accretion theory), our results may be used to constrain $Q/10^6\lesssim M_c/(6M_\oplus)$. Notice, however, that a high $Q>10^7$ is also consistent with our findings, because in that case most of the Jupiters in Fig. \ref{fig:ator} will not inspiral towards the Roche limit, according to Equation \eqref{eq:t_tide}. In this scenario, the lack of observed RLO remnants merely reflects the scarcity of their possible progenitors.  
\item
The planet's evolution during RLO is essentially a reflection of its density $\rho_p(M_p)$. Therefore, a more accurate calculation, involving a sophisticated equation of state, can improve our constraint (see, however, Section \ref{sec:previous}).
\item\label{item:pe}
While gas Neptunes are unaffected by photo-evaporation (PE), lighter planets are more vulnerable. Such super Earths, which may be the result of RLO with low $Q$ values, necessitate an analysis of PE during their late evolution. In addition, it is crucial to reconcile the difference between the PE prescriptions of \citet{LopezFortney2013} and \citet{Valsecchi2015} at late times.
\end{enumerate}

In summary, by identifying the (missing) remnants of inspiralling hot Jupiters that undergo Roche-lobe overflow, we were able to constrain their core mass. This constraint may be used to distinguish between different formation scenarios of gas giants. Moreover, our analysis shows that short-period low-density super Earths cannot be the result of RLO of inspiralling Jupiters, unless $Q$ is low. Instead, these planets could have formed by core-accretion that did not reach the runaway phase, as their atmosphere mass fractions $M_{\rm atm}/M_c\lesssim 0.5$ suggest \citep[see, e.g.,][and references therein]{Ginzburg2016,LeeChiang2016}. 

\section*{Acknowledgements}

This research was partially supported by ISF (Israel Science Foundation) and iCore (Israeli Centers of Research Excellence) grants.
This research has made use of the Exoplanet Orbit Database and the Exoplanet Data Explorer at exoplanets.org.
We thank Kenta Hotokezaka, Itai Linial, and Brian Metzger for discussions and comments and Brian Jackson for a comprehensive review that improved the paper.




\bibliographystyle{mnras}
\bibliography{roche} 



\appendix
\section{Overflow Stability}\label{sec:stability}

The stability and rate of the Roche-lobe overflow, once the Roche limit is reached and the planet starts to lose mass ($dM_p<0$), is based on \citet{Rappaport1982}, with a brief description here \citep[see also][]{Valsecchi2014,Valsecchi2015,Jackson2016}.

In the absence of tides, the orbital angular momentum $L=M_p\sqrt{GM_\star a}$ is conserved (though see the discussion below) during mass transfer, so that $d\ln a=-2d\ln M_p$. The expanding orbit and the shrinking mass change the planet's Roche-lobe radius $R_L\sim a(M_p/M_\star)^{1/3}$. Specifically, $d\ln R_L=d\ln a+(1/3)d\ln M_p=-(5/3)d\ln M_p$. The response of the planet's radius to the mass loss is given by its mass-radius relation, parametrized by $\xi\equiv d\ln R_p/d\ln M_p$. When the planet reaches the Roche limit, $R_p=R_L$ by definition, as realized in Equation \eqref{eq:roche}. Therefore, the sign of $d\ln R_L-d\ln R_p=-d\ln M_p(5/3+\xi)$ determines whether the mass transfer causes the planet to overflow its Roche lobe even more (triggering unstable mass transfer) or whether it shuts down the overflow. For gas giants $\xi>-1/3$ (see Section \ref{sec:compress}), promoting stable mass transfer. When tides, or any other orbital angular momentum loss mechanism (e.g. gravitational waves) are included, the stable mass transfer ensures that the planet is kept on the verge of overflow, so that $R_p=R_L$, and Equation \eqref{eq:roche} is satisfied. Moreover, by adding angular momentum loss to the analysis above, we find that $2d\ln L=(5/3+\xi)d\ln M_p$, implying that the Roche-lobe overflow time-scale is determined by Equation \eqref{eq:t_tide}. 

These results are easily generalized for a non-conservative mass transfer scenario, in which a fraction of the mass escapes from the star-planet system with a fraction of its orbital angular momentum $d\ln L_{\rm mass-loss}\equiv\alpha d\ln M_p$. In this case, the mass-transfer rate is given by
\begin{equation}\label{eq:rlo_rate}
\frac{|\dot{M_p}|}{M_p}=\left(\frac{2}{5/3+\xi-2\alpha}\right)t_{\rm tide}^{-1},
\end{equation}
where the rate of orbital angular momentum loss by tides is given by Equation \eqref{eq:t_tide}, and where the stability criterion is $5/3+\xi-2\alpha>0$. As noted by \citet{Metzger2012} and \citet{Jackson2016}, even if no mass is lost from the system, $\alpha$ can be as large as $\sqrt{R_\star/a}$ because the accreted mass can transfer orbital angular momentum to the spin of the star. Using Equation \eqref{eq:roche} with the coincidence $\rho_p\approx\rho_\star$ (see Section \ref{sec:inspiral}) and $\xi\approx 1/3$ (for a roughly constant density, which is adequate for $M_c<M_p<3M_{\rm J}$, as explained in Section \ref{sec:rlo}), we conclude that the mass transfer is stable if less than about a third of the mass escapes the system (or if it does not carry with it its entire original angular momentum). When $M_p\sim M_c$, the density and therefore $\sqrt{R_\star/a}$ increase (see Sections \ref{sec:rlo} and \ref{sec:mc}), but $\xi\approx M_p/(3M_{\rm atm})$ increases more dramatically, so the mass transfer remains stable.

To summarize, the Roche-lobe overflow is stable, unless the mass transfer if highly non-conservative. Determining the amount of mass and angular momentum loss from the system is beyond the scope of this work \citep[see, e.g.,][]{Itai}, so we follow previous studies and assume a stable overflow.


\bsp	
\label{lastpage}
\end{document}